# Object Detection for Caries or Pit and Fissure Sealing Requirement in Children's First Permanent Molars

Chenyao Jiang[a,b,†], Shiyao Zhai[a,b,†], Hengrui Song[c], Yuqing Ma[a,b], Yachen Fan[a,b], Yancheng Fang[d], Dongmei Yu[e], Canyang Zhang[b], Sanyang Han[b], Runming Wang[b], Yong Liu[f], Jianbo Li[g,*], and Peiwu Qin[a,b,*]

Corresponding Author

* Peiwu Qin, E-mail: pwqin@sz.tsinghua.edu.cn.

* Jianbo Li, E-mail: 435264691@qq.com.

[a] Center of Precision Medicine and Healthcare, Tsinghua-Berkeley Shenzhen Institute, Shenzhen, Guangdong Province, 518055, China

[b] Institute of Biopharmaceutical and Health Engineering, Tsinghua Shenzhen International Graduate School, Shenzhen, Guangdong Province, 518055, China

[c] Division of Information Science and Technology, Tsinghua Shenzhen International Graduate School, Shenzhen, Guangdong Province, 518055, China

[d] Shenzhen Stomatology Hospital (Pingshan), Southern Medical University, Shenzhen, Guangdong Province, 518055, China

[e] School of Mechanical, Electrical & Information Engineering, Shandong University, Weihai, Shandong, 264209, China.

[f] Nanshan Center for Chronic Disease Control, Shenzhen, Guangdong Province, 518055, China

[g] Department of Preventive Dentistry, Stomatological Hospital, Southern Medical University, Guangzhou, Guangdong Province, 510115, China

**Abstract**

Dental caries is one of the most common oral diseases that, if left untreated, can lead to a variety of oral problems. It mainly occurs inside the pits and fissures on the occlusal/buccal/palatal surfaces of molars and children are a high-risk group for pit and fissure caries in permanent molars. Pit and fissure sealing is one of the most effective methods that is widely used in prevention of pit and fissure caries. However, current detection of pits and fissures or caries depends primarily on the experienced dentists, which ordinary parents do not have, and children may miss the remedial treatment without timely detection. To address this issue, we present a method to autodetect caries and pit and fissure sealing requirements using oral photos taken by smartphones. We use the YOLOv5 and YOLOX models and adopt a tiling strategy to reduce information loss during image pre-processing. The best result for YOLOXs model with tiling strategy is 72.3 mAP.5, while the best result without tiling strategy is 71.2. YOLOv5s6 model with/without tiling attains 70.9/67.9 mAP.5, respectively. We deploy the pre-trained network to mobile devices as a WeChat applet, allowing in-home detection by parents or children guardian.

## 1. Introduction

Dental caries is a multi-factorial chronic oral disease that places a major oral health burden for many countries[1]. According to the Global Burden of Disease 2019, dental caries of primary teeth affects over 520 million children[2], and untreated dental caries in permanent teeth is the most frequent health problem[1,3,4]. Dental caries can lead to pulpitis, periapical infection, and even tooth loss if not treated properly[5]. The period of highest risk of carious lesion development in permanent teeth is the first few years after tooth eruption[6]. Despite the fact that great efforts to prevent caries have been implemented for children and adolescents, the pit and fissure caries (PFC) among young people is more prevalent than caries on smooth surfaces[7].

Caries is most commonly found inside the pits and fissures on the occlusal/buccal/palatal surfaces of the molars. PFC accounts for about 90% of the caries in permanent molars and 44% of the caries in the deciduous molars of children and adolescents[8]. Early caries detection and prevention can effectively halt the further deterioration of caries while reducing the symptom and economic burden. Brushing with fluoride toothpaste, fluoride supplements, local enamel fluoridation, and pit and fissure sealing (PFS) are available methods for preventing dental caries[9]. PFS is a preventive conservation approach that involves the placement of sealants into the pits and fissures of caries prone teeth. Sealant is applied to the surface of the molars to create a physical barrier and prevent bacterial growth, which can lead to dental caries, by blocking nutrition. Caries prevention by PFS in children and adolescents is well established and studies show that sealant reduces caries increment by 11% to 51% over a two-year follow-up study, compared with no sealant applied to the first permanent molars[10]. Sealants are not only regarded as an effective material for preventing caries on occlusal surfaces, but also considered as the active agent in controlling and managing initial caries lesions on occlusal and approximal surfaces[11].

Rapid, timely, and precise detection and diagnosis are vital for implementing dental sealants. Clinically, visual-tactile detection of pit and fissure or caries is the most common method[12].

Although the visual-tactile method depends more on the clinically experienced dentists, it avoids the high cost and complicated imaging, and remains the most direct and common clinical judgment method, compared with other diagnostic aids. Due to the global oral health resource imbalance, we must admit that despite significant investment in caries prevention for children and adolescents, many families are still unable to detect the caries in time and miss the best time range to perform PFS on their children, resulting in irreversible damage to permanent molars. Therefore, it is essential to develop a simple and convenient method to detect kids' caries by parents themselves. Additionally, this method can add more protection for children during times of epidemic outbreaks, such as COVID-19, flu and monkeypox[13], by allowing in-home detection for their teeth.

Object detection is one fundamental task of computer vision and deep learning has recently demonstrated excellent performance in this area[14]. The goal of object detection is to locate and classify object category in an image or video. Deep models for object detection can be divided into two categories: (1) two-stage methods, such as the R-CNN families (Fast R-CNN[15], Mask R-CNN[16],) and their variants (FCOS[17], PAA[18] and ATSS[19]). This kind of method first locates the potential target, obtains a number of ROIs, and then identifies each ROI. (2) one-stage methods. The YOLO series, RetinaNet, and M2Det serve as examples of this type of methodology. One-stage methods complete the two tasks of positioning and classification at once, and thus has advantages in speed, while the two-stage models typically have better positioning and recognition accuracy. The dataset used to train the model has a impact on how well deep models perform. The representative of commonly used object detection datasets includes PASCAL VOC[20], MS COCO[21], ImageNet[22] and Object365[23]. PASCAL VOC is an early published dataset (in 2010) and contains over 10,000 images of objects in 20 different categories that are marked by bounding boxes. While MS COCO, ImageNet and Object365 are larger datasets, the number of images in these datasets are 330k, 500k and 630k, respectively, and that of object categories is 80, 200 and 365. The success of vision transformers (ViT)[24] brings another popular choice for object detection task, since the large-scale vision foundation models can be trained using self-supervised tasks on unlabeled data and then be fine-tuned on downstream tasks like classification, segmentation, detection, and multi-modal tasks. Even though the pretrained models shows great capacity on benchmark datasets, there are still many challenges when applying them to specific domains. First, the images in previous mentioned datasets are usually realistic scene images which may differ greatly from those seen in real application such as detecting in X-ray images, and the performance of the pretrained models may reduce when features are extracted from a different data distribution[25]. Model tuning presents another difficulty. Due to the high cost of labeling and inadequate data source, the datasets of specific domain have a relatively small size, which limits the effectiveness of deep models. In addition, there are problems such as large image size, small object, dark light, occlusion and complex environments. When the image size is large and the objects to be detected are small, after being reshaped to a smaller size (in order to be fed in to the deep network), there will be less information for the target objects, which will cause the difficulty in detection. For small objects that take up little space in the image, there is less feature information that can be used to distinguish them. In addition, the IoU will be low if there is only a slight deviation between the predict bounding box and the ground truth label, which is also the cause of the decline in network performance. It is a difficult problem to perform the detection task in low light condition. Images of dark condition do not have sufficient features for visual processing. The long exposure time could lead to noise and motion blur that affects visual tasks. Occluded targets in datasets such as COCO are often presented as hard cases. Because of occlusion, some features of the target object are invisible in the image, which causes difficulties in detection. The state-of-the-art models may not perform well on tasks under certain circumstance, thus it is important to use some strategies based on the specific difficulty of the dataset.

Deep models have been adopted for medical diagnosis in a variety of clinical dental settings[13,26-32]. Deep convolutional neural networks have been applied to periapical radiographs and near-infrared transillumination images for dental caries detection and diagnosis, demonstrating that deep learning approaches for analysis of dental images hold promise for increasing the speed and accuracy of caries detection, supporting the dental practitioners' diagnoses[33] [34] [35,36]. The Mask R-CNN detects and classifies dental caries on occlusal surfaces across the entire 7-class on an intraoral camera image dataset[37]. With the popularity and advancement of smartphones, there are numerous opportunities to predict caries using images taken by cell phones, which are more convenient and ubiquitous. Zhang et al. developed a model to classify and locate the existence of dental caries with bounding boxes from oral images taken by smartphone[38]. However, the use of a faster target detection network for PFS determination rather than just caries detection may provide more clinical assistance.

In this work, we train YOLOv5 and YOLOX models and deploy them on the cloud server to achieve a real-time and fast detection for children's caries and PFS requirements in permanent molars using images taken by cell phones and uploaded by the users. To the best of our knowledge, there is no available datasets for such a purpose. We collect a dataset of children's oral cavity taken by cell phones. Our dataset consists of 4563 oral photographs taken by cell phones for children aged seven to nine years, which is in the highest risk period involved by PFC of the first permanent molars. The detection task on our dataset suffers from the problem of small target object and complex environment, so we use a tilling and a merge strategy before and after detection with YOLOv5 and YOLOX model. As for the result, the YOLOX with tiling method achieves 72.3% mean average precision (mAP), which is the best of our tested models and methods. In order to improve its mobile usefulness in real life, we choose WeChat applet as the front end since the widespread use of WeChat can reduce the difficulty of promotion and the operation threshold, and deploy the pretrained model on the cloud server. This method brings not only lower testing costs but also more frequent detections for children to avoid missing the PFS due to its flexibility and convenience. More home dental screenings for children will open up the possibility of more equitable and convenient medical resource sharing and reduce disease treatment burden.

## 2. Materials and Methods
### 2.1 Dataset
The raw dataset has 16023 oral images taken by cell phones, and is provided by a hospital from Guangzhou, China. The volunteers involved range in age from 7 to 9 years old, and are in the first few years after the eruption of the first permanent molars, when PFC are common. Dental experts use Labelme software to label the first permanent molars in pictures and classify them into three categories: caries, PFS requirement and no PFS requirement, as shown in **Figure 1**. Some images contain multiple instances (**Figure 1D**). The PFS-required molars should meet three criteria: full eruption, no caries, and deep pit and fissure. Permanent molars that have not fully erupted or with very shallow pit and fissure fall into the no PFS requirement category. The oral photographs are taken by cell phones in a realistic environment. In our dataset, (1) the images in the dataset are not uniform in size, which ranges from ~500x500 to ~3000x4000. (2) The images are of various resolution, and in low-resolution images, the surface of teeth is not that clear. (3) Additionally, the brightness or light condition of the oral cavity in some images may be insufficient. (4)Finally, some images cannot fully show the surface of the permanent molars due to the shooting angle, or the permanent molars are partially obscured by other teeth. We remove images that are duplicates or difficult for dental professionals to identify, but the above problems may still exist in remaining images (mainly the first three kinds of problem). The final dataset consists of 4563 oral pictures (5077 instances). The dataset is randomly divided into a training set (60%), a validation set (20%), and a test set (20%) (**Table 1**). We can provide the dataset and label files in YOLO and COCO versions upon request.

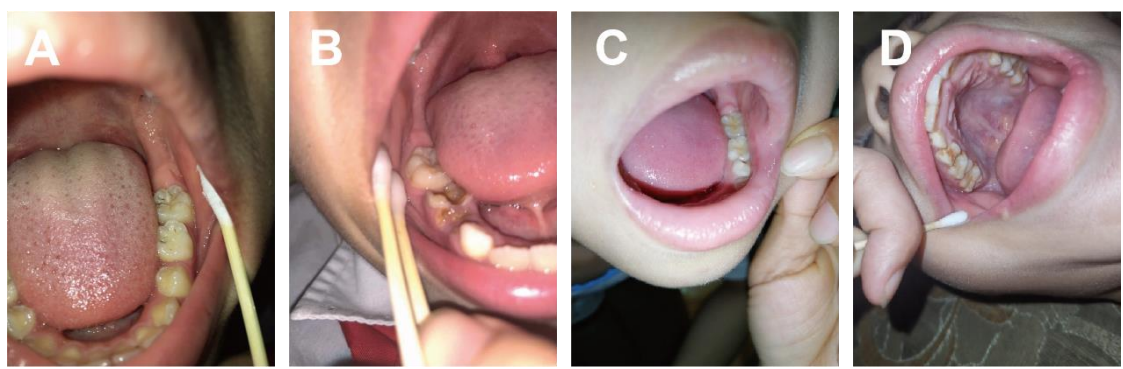

**Figure 1** The example images in the dataset. All these images contain the first permanent molars. A. Caries; B. PFS requirement; C. no PFS requirement; D. contains multiple instances.

Table 1 Details of dividing dataset.

|  | Caries | PFS Requirement | No PFS Requirement | All |
|---|---|---|---|---|
| **Train set instances** | 1273 | 617 | 1143 | 3033 |
| **Validation set instances** | 444 | 210 | 373 | 1027 |
| **Test set instances** | 439 | 220 | 394 | 1053 |
| **All** | 2156 | 1047 | 1910 | 5077 |

## 2.2 The Object Detection Algorithms

We select YOLOV5 (*https://github.com/ultralytics/yolov5*) and YOLOX (*https://github.com/Megvii-BaseDetection/YOLOX*) for the target detection of dental caries and PFS requirements. Their basic structures are similar to the structure of YOLOV3[39] and can be divided into four parts: input, skeleton, bottleneck and head. Both of the models contain different network versions, ranging from small to large, implemented by varying the model depth and width: YOLOV5s/YOLOX-s, YOLOV5m/YOLOX-m, YOLOV5l/YOLOX-l, YOLOV5x/YOLOX-x. Accuracy on benchmark dataset will improve in larger networks, but the number of parameters, training costs, and inference time will grow as well.

YOLOV5 conducts mosaic data enhancement, adaptive anchor frame calculation and adaptive image scaling to the input image. The added Cross Stage Partial (CSP) is used as a backbone focusing on the feature extraction in low processing time. YOLOV5 includes a bottom-up feature pyramid structure on top of the Feature Pyramid Network (FPN) structure. Through this combination operation, the FPN layer conveys strong semantic features from the top down and the feature pyramid delivers strong localization features from the bottom up. The feature aggregation of different feature layers improves the network's ability to detect targets at different scales. The head part performs final detection in seed image, calculating bounding boxes by applying anchor boxes on features.

Compared with YOLOV3[39], YOLOX uses DarkNet53 as its backbone and introduces Mosaic and modified MixUp for data augmentation. The main differences between YOLOX and YOLO series models are as follows: (1) Anchor free. The clustering result on the dataset determines the size and position of the anchor boxes in the anchor mechanism, increasing the model's complexity. YOLOX achieves comparable results while reducing training parameters. (2) Decoupling head. Compared with other YOLO models' single head, YOLOX employs a decoupling head with two branches: one classification branch with a head for classification and another regression branch with a head for calculating bounding box positions and a head for confidence. The decoupling head converges faster than other YOLO's single head.

**2.3 Tiling And Tile-Filtering**
Because of the non-professional oral equipment and photographic methods, the images of datasets collected at home usually have a complex background around the oral cavity, including other parts of the human face, clothing, hair, etc. In some images, the background occupies the majority of the image, leaving only a small portion for the oral cavity, and the permanent molars and fissures to be detected are even smaller. In the target detection task, due to hardware limitations and detection speed requirements, the detected image is usually resized to a smaller size before being inputted into the network, leading to a loss of the information about the surface of permanent molars during the resize process. During the prediction process, the localization of small targets is more challenging compared with large/medium targets because a small target covers fewer area in the image and the offset of the prediction frame has a greater impact on the error. Therefore, the target detection model has a discrepancy in the detection accuracy for large and small targets. When the image size is 640 × 640, the AP of YOLOX-M/L/X on the small target of the COCO2017 dataset is 26.3/29.8/31.2, which is significantly lower than 51.0/54.5/56.1 on medium targets or 59.9/64.4/66.1 on large targets.

In order to alleviate detection problems for small object, we adopt a tiling strategy during model training and inference[40]. During the training process, each image in the training set will be divided into N × M overlapping tiles, which together with the full image form an extended dataset to train the network, as shown in **Figure 2**. Due to the extensive background in the image, not every tile contains the teeth to be detected; abundant background information may interfere with detection. As a result, we develop a tile filter that selectively allows inputting the oral cavity tiles into the detection network, while those containing only background information are discarded.

To train a tile filter, we first build a new dataset. We crop images from the training and validation sets into tiles using the tiling strategy. The tiles are labeled as positive or negative based on whether they contain a target frame. Since the number of negative samples is much greater than the number of positive samples, we choose the negative samples at random and the final ratio of positive to negative samples is 1:1. As for the feature to classify the tiles, the common choices are traditional image features, such as histogram, and deep-model-based feature, for example, the feature output from a pretrained VGG16 network. Considering the following three factors we choose the histogram as the depictor of the tile: (1) the difference between colors of oral cavity and the background area is relatively obvious, there may be no need to use a deep learning model to extract the feature from the tile. (2) The histogram as feature to describe an image is a one-dimensional array and is fast to generate, and the deep model is time-consuming, which is not a good way to meet the goal to build a real-time detector. (3) The accuracy of the classifier based on histogram as the image feature is satisfied enough (see in result). Finally, the tile filter will classify the histogram array of images. The tiling and tile-filter methods are exploited during the inference process.

After tilling and tile-filtering, each tile is processed independently as well as the original image. The resulting detection boxes with class probabilities are collected as the initial results. In the merging step, the full image results and the tile results are combined to form the final output.

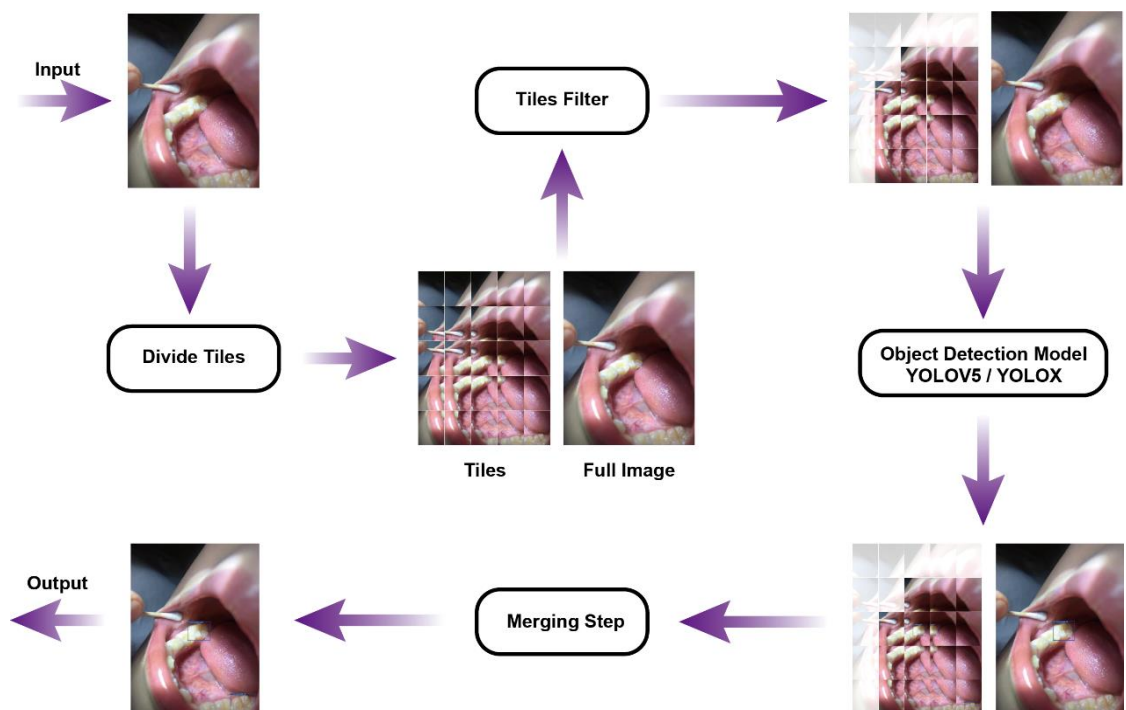

**Figure 2** The workflow of tile and tile-filter strategy and merging step in the analysis pipeline.

### 2.4 Metric

In the object detection task, mean Average Precision (mAP) is the main evaluation metric that reflects the global performance. For each category, we calculate the Precision and Recall firstly. Precision measures the accuracy of predictions, and Recall indicates the performance in finding all positives. mAP is the mean value of APs in all categories, and AP could be obtained by calculating the area under the corresponding Precision-Recall curve using the 11-point interpolation technique introduced in the PASCAL VOC challenge. mAP@.5 represents the average AP when the intersection over union (IoU), positive threshold, is 0.5. mAP@.5:.95 corresponds to the average

AP for IoU from 0.5 to 0.95 with a step size of 0.05. The mathematical definitions are shown in the following formulas.

$$Precision = \frac{TP}{TP + FP}$$

$$Recall = \frac{TP}{TP + FN}$$

$$AP = \frac{1}{11} \sum_{recall \in [0, 0.1, \ldots, 1]} Precision(Recall)$$

$$IoU = \frac{area\ of\ overlap}{area\ of\ union}$$

$$mAP = \frac{1}{c} \sum_{i=1}^{c} AP_i$$

Where the abbreviations are listed as follows: True Positive (TP); False Positive (FP); False Negative (FN); categories (c).

**2.5 Applet Design**

Considering the practical application to home prescreening for children's teeth, we deploy the detection method to the mobile. We make a WeChat applet demo for mobile deployment because the applet requires no additional download and is easy to promote and operate. The front-end is built with the WeChat developer tool and the Flask framework is used to package the tiling strategy and the pretrained model, as well as design the Application Programming Interfaces (APIs) on the Tencent cloud server. As shown in **Figure 3**, the user can use the front-end to take the first permanent molar pictures under detection and upload them to the server through the post API. The server employs the tiling and tile-filtering strategies to preprocess photos. Immediately, the trained network completes the task of object detection of the picture and the merging strategy combines detection results from the full image and tiles. The final outputs are temporarily saved. When the front-end initiates a request to view the results, the server will send the predicted results to the user's cell phone, providing the pictures of the first permanent molar target detection and labeling.

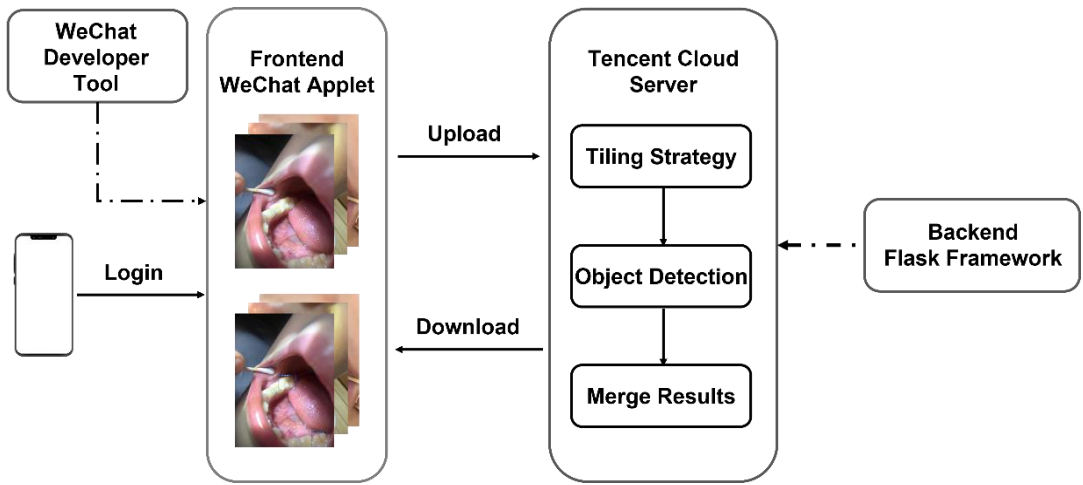

**Figure 3** The frontend and backend design of WeChat applets and their communication.

### 2.6 Experimental details

In the tiling process, all images are tiled into $5 \times 5$ grids (i.e., N = 5 and M = 5) with the help of overlapping tiles, and the intersection between consecutive tiles is 50%. We select the Support Vector Machine (SVM) as the tile filter. We finalize 5600 tile images (2800 target tiles and 2800 nontarget tiles) as the training set and 1700 tile images (850 target tiles and 850 nontarget tiles) as the test set for the dataset used to train the tile filter. In the classification task, the SVM classifier achieves 77.3 percent accuracy and 90.5 percent recall.

We select YOLOv5s6 and YOLOXs as the object detection models, considering our computation resource and the need for quick detection. For training, the image size is $640 \times 640$. There are two versions of the training set: one with full images ($D_f$) and one with both full and tile images ($D_{ft}$). Since both models are trained on the two dataset versions, there are a total of four combinations: (1) YOLOv5s6 trained on $D_f$ (referred to as YOLOv5), (2) Yolov5s6 trained on $D_{ft}$ (referred to as YOLOv5-tiling), (3) YOLOXs trained on $D_f$ (referred to as YOLOX), (4) YOLOXs trained on $D_{ft}$ (referred to as YOLOX-tiling). The experiment settings (learning rate, weight decay) of models are similar to the original implementations of YOLOv5 and YOLOX , with the exception of data augmentation. The probability of applying mosaic augmentation is changed to 0.5 and the MixUp augmentation is 0 because according to our experiments, the model will reach a higher mAP when the parameters are modefied in this way. All the models are trained 300 epochs, and the best checkpoint of each model is used in testing.

During testing, for YOLOv5 and YOLOX, the size of the input images is $640 \times 640$. For YOLOv5-tiling and YOLOX-tiling, the input images are firstly cropped to $5 \times 5$ tiles using tiling strategy. After removing nontarget tiles with a tile-filter, the remaining target tiles and the full image will be resized to $640 \times 640$ to serve as the model input. The bounding boxes generated by tiles and the full image perform non-maximum suppression (NMS) jointly to obtain the output results. The NMS threshold is set as 0.45 for all models.

## 3. Results and Discussion
### 3.1 Object Detection with YOLOv5 and YOLOX

On the test set, YOLOv5 and YOLOX achieve 67.9 and 71.2 mAP.5 respectively (Table 2). Caries detection works best among the three detection tasks: YOLOv5 and YOLOX achieving 79.7 and 84.7 mAP.5 on No_PFS_Requirement/PFS_Requirement, YOLOv5 attains 62.7/61.2 mAP.5 and YOLOX is 62.9/66.5 mAP.5. We attribute this result to the more obvious caries characteristics, whereas the depth of the pits and fissures are subtle and difficult-to-detect features.

**Table 2** The performance of the four models in caries and PFS object detection.

| Model | mAP (%) | | | | | | | |
| --- | --- | --- | --- | --- | --- | --- | --- | --- |
| | No PFS Requirement | | PFS Requirement | | Caries | | All | |
| | 0.5:0.95 | 0.5 | 0.5:0.95 | 0.5 | 0.5:0.95 | 0.5 | 0.5:0.95 | 0.5 |
| YOLOv5 | 25.2 | 62.7 | 30.0 | 61.2 | 38.1 | 79.7 | 31.1 | 67.9 |

| | | | | | | | |
|---|---|---|---|---|---|---|---|
| YOLOv5-Tiling | 25.9 | 64.9 | 31.1 | 65.9 | 38.6 | 81.8 | 31.9 | 70.9 |
| YOLOX | 25.4 | 62.9 | 32.9 | 66.5 | 39.9 | 84.7 | 32.7 | 71.2 |
| YOLOX-Tiling | 26.1 | 64.2 | 33.2 | 67.1 | 40.1 | 85.6 | 33.1 | 72.3 |

The tiling strategy improves the model performance. YOLOv5s6-Tiling reaches 70.9 mAP.5 in the test set, and YOLOXs-Tiling achieves 72.3 mAP.5 as the best result. We visualize the detection results (**Figure 4**), compare the model outputs with and without the tiling strategy, and attribute the gain brought by tiling to the three reasons listed below:

(1) Tiling models tend to detect all permanent molars, whereas non-tiling models may miss some. In the figure, the non-tiling model sometimes ignores the permanent molar in the lower right corner (**Figure 4A**), but the tiling model detects it correctly. This is because the tiling strategy avoids information loss during the resizing process, allowing the model to detect it better.

(2) Based on image details, the tiling model can make more accurate decisions. The non-tiling model classifies the permanent molars on the left side of some images that require PFS as not requiring PFS, but the tiling model provides the correct classification result (**Figure 4B**). This is also a result of more retained information from the image.

(3) The tiling model produces more statistically confident results. As shown in **Figure 4C**, both the non-tiling model and the tiling model correctly detect and classify the two caries in the picture, with the tiling model having higher confidence levels (78.3% and 73.7%) than the non-tiling model (73.6% and 69.9%).

However, the results also reveal certain issues caused by the tiling strategy. Our goal is to detect the first permanent molars, which are the innermost teeth in the oral cavity when children are 7 to 9 years old. Determining the position of the permanent molars in the image requires information of the entire oral cavity. Nevertheless, some tiles only show parts of the oral cavity, making it difficult to tell whether the teeth in these tiles are permanent molars based on a single tile. As shown in **Figure 4D**, the tiling model may recognize it as other kinds of teeth during the detection process, whereas the non-tiling model does not.

As mentioned in section 2.1, the problems we face when do object detection on our dataset are small objects, low resolution, and low light situation.

In addition to tilling strategy, the problem of small target detection or low resolution can be solved by improving image resolution (super resolution) in preprocessing stage. Super resolution can restore more details of the image, thus improving the detection result. However, the super resolution step will result in a larger size of image, which will increase the detection time. Therefore, there is a trade-off between detection efficiency and accuracy. We use BSRGAN[41] for a 2x super-resolution preprocessing. BSRGAN is a single image super resolution model trained with images generated by a degradation model that consists of blur, noise and downsampling, and covers a wide range of image degradations found in real-word scenarios. We use the pretrained weights of BSRGAN instead of training a new model on our dataset because we do not have enough high resolution images for super resolution training. Through our experiments, the YOLOX model with tilling strategy and a super-resolution preprocessing step gain an mAP of 71.9 when testing, which is lower than the model only with a tilling step. This result shows that the super resolution strategy is unable to further improve the detection performance in our task. This might be because (1) we use the model weights of BSRGAN trained on other realictic scene datasets which has a different data distribution with ours, thus the super-resolution model cannot well restore the details of teeth, and

(2) context information contains in a tile will become less with the 2x super-resolution step which doubled the image size while the tile size remains the same. In addition, super resolution required more time and computing resource, so we abandoned this strategy.

The low light problem can be solved through the data augmentation method which is already involved in the training process of YOLOv5 and YOLOX. The V channel in the HSV (Hue, Saturation and Value) color space corresponds to the brightness of the image. HSV augmentation add disturbance to the three channels, and changes on the V channel can change the brightness of image during training, making models generalize well to images of different light conditions.

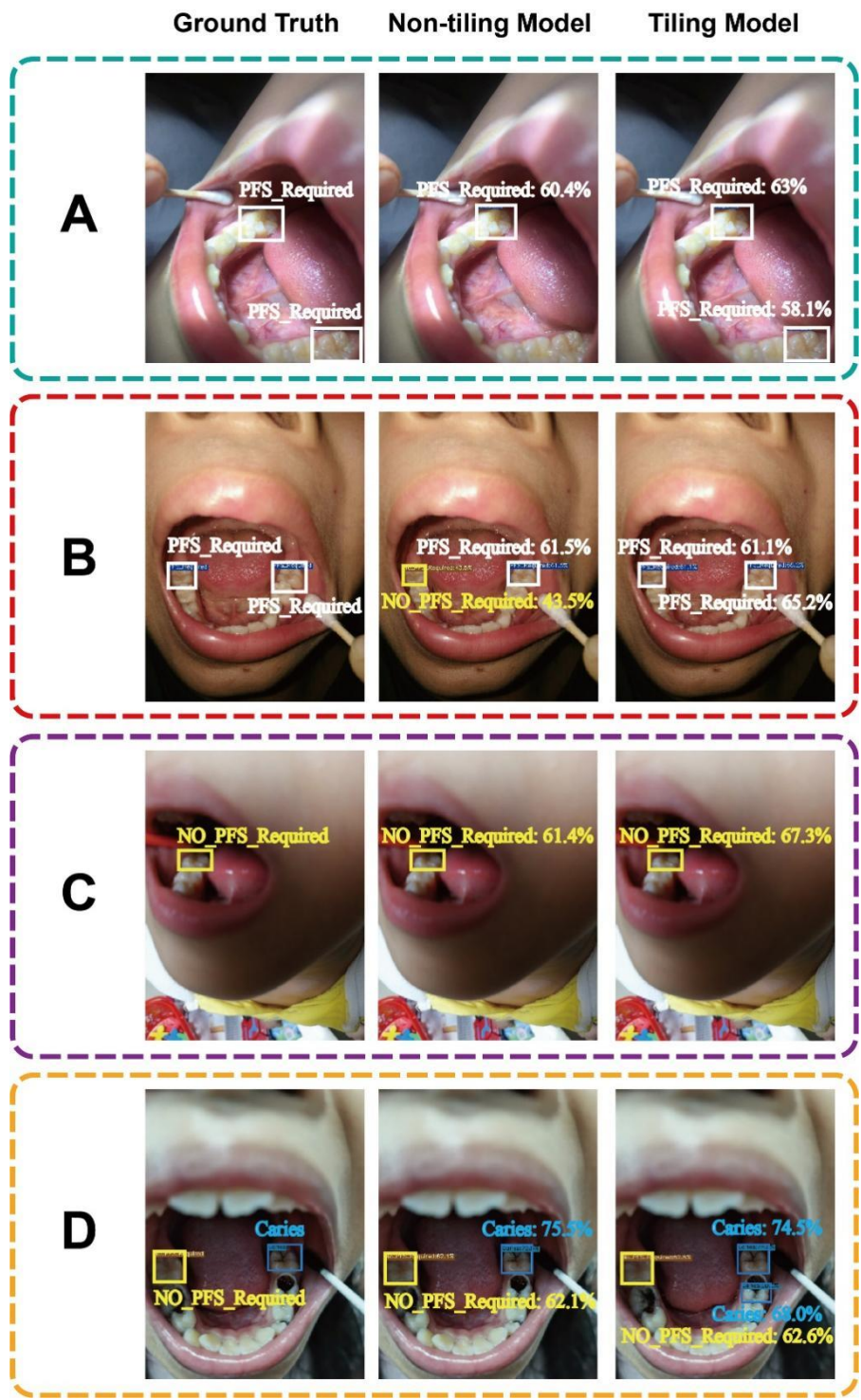

**Figure 4** Visualization results from the ground truth, the non-tilling model, and the tilling model. Compared with non-tilling model, the tilling model shows better performance (A-C) although it may sometimes detect caries in teeth that are not the first permanent molars (D).

### 3.2 WeChat Applet for Caries and PFS Object Detection

The detection network deployed in the applet is YOLOXs, which has the best inference performance. Users can login the applet directly without any cumbersome operations. As shown in **Figure 5**, the tidy front-end interface provides users with "Image Upload" and "Result Download" buttons. The "Image Upload" function connects to a post API on server and supports the actions of waking up the camera and posting pictures, or directly uploading local photos. The "Result Download" function allows downloading the predicted results from the server for local storage. It shows the bounding boxes that label the first permanent molars and provides the prediction labels with confidence.

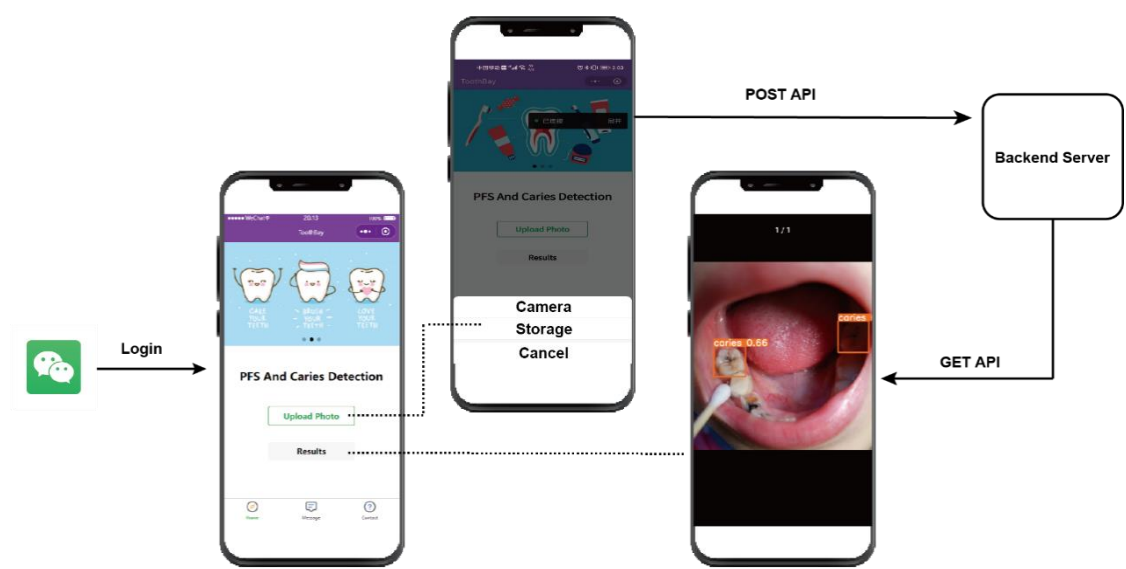

**Figure 5** The WeChat applet demo for PFS and caries object detection.

### 4. Conclusion

PFS is one of the most effective methods to prevent PFC which is the main dental caries in children and adolescents. Timely and accurate detection can effectively prevent children from missing the optimal time period to have PFS. In this work, we demonstrate the usage of deep learning networks YOLOV5 and YOLOX to achieve object detection for PFS requirement and caries of first permanent molars in children. We introduce tiling and tile-filter strategies to improve the network performance for small target detection and obtain 72.3% mAP.5. In addition, we deploy the pre-trained network to mobile devices as an applet. Our detection does not require complex processes or professional dental equipment, and it will really help achieve a home pre-screening for children's PFS needs. In the future, the model's performance is expected to improve further after being trained on the image data collected using the applet. More image data could help explore solutions for detecting small objects in dark and low-resolution images. This model and Applet can be used for screening of total oral caries, PFS requirements, or other oral diseases as well. Furthermore, Neuro-Symbolic Learning (NeSyL)[42], which combines symbolic artificial intelligence and connectionist artificial intelligence, can overcome the problem of data distribution imbalance in our dataset. It can also be utilized to improve the model's interpretability by incorporating a priori knowledge of dentistry, resulting in providing a more trustworthy diagnostic basis for home screening.

**Acknowledgements**

This work was supported in part by Science, Technology, Innovation Commission of Shenzhen Municipality (JSGG20191129110812708; JSGG20200225150707332; ZDSYS202008201654000; JCYJ20190809180003689; WDZC20200820173710001), National Natural Science Foundation of China (31970752), and Shenzhen Bay Laboratory Open Funding (SZBL2020090501004).

**Declaration of Interests**

The authors declare no conflict of interests.